\documentclass[conference]{IEEEtran}
\IEEEoverridecommandlockouts

\usepackage{cite}
\usepackage{amsmath,amssymb,amsfonts}
\usepackage{algorithmic}
\usepackage{graphicx}
\usepackage{textcomp}
\usepackage{xcolor}
\def\BibTeX{{\rm B\kern-.05em{\sc i\kern-.025em b}\kern-.08em
    T\kern-.1667em\lower.7ex\hbox{E}\kern-.125emX}}
\begin{document}

\title{Towards Expressive Video Dubbing with Multiscale Multimodal Context Interaction\\
}

\author{\IEEEauthorblockN{Yuan Zhao}
\IEEEauthorblockA{\textit{Inner Mongolia University} \\
Hohhot, China \\
zy404nf@163.com}
\and
\IEEEauthorblockN{Rui Liu*\thanks{* Corresponding Author.}}
\IEEEauthorblockA{\textit{Inner Mongolia University} \\
Hohhot, China \\
liurui\_imu@163.com}
\and
\IEEEauthorblockN{Gaoxiang Cong}
\IEEEauthorblockA{\textit{Inst. of Comput. Tech., CAS} \\
Beijing, China \\
conggaoxiang24b@ict.ac.cn}


}


\maketitle

\begin{abstract}
Automatic Video Dubbing (AVD) generates speech aligned with lip motion and facial emotion from scripts. Recent research focuses on modeling multimodal context to enhance prosody expressiveness but overlooks two key issues: 1) Multiscale prosody expression attributes in the context influence the current sentence's prosody. 2) Prosody cues in context interact with the current sentence, impacting the final prosody expressiveness. To tackle these challenges, we propose M2CI-Dubber, a Multiscale Multimodal Context Interaction scheme for AVD. This scheme includes two shared M2CI encoders to model the multiscale multimodal context and facilitate its deep interaction with the current sentence. By extracting global and local features for each modality in the context, utilizing attention-based mechanisms for aggregation and interaction, and employing an interaction-based graph attention network for fusion, the proposed approach enhances the prosody  expressiveness of synthesized speech for the current sentence.
Experiments on the Chem dataset show our model outperforms baselines in dubbing expressiveness. The code and demos are available at \textcolor[rgb]{0.93,0.0,0.47}{https://github.com/AI-S2-Lab/M2CI-Dubber}.
\end{abstract}

\begin{IEEEkeywords}
Automatic Video Dubbing, Multiscale Multimodal Context Interaction, Prosody Expressiveness.
\end{IEEEkeywords}

\vspace{-3mm}
\section{Introduction}
\vspace{-1mm}
Video dubbing is used to re-record audio originally captured in noisy environments, resulting in unusable quality \cite{hu2021neural}. However, video dubbing is expensive, as it requires studios, professional actors, and considerable recording time \cite{3D-VD}. Automatic Video Dubbing (AVD) aims to lower these costs by generating speech from scripts of the given current sentence while ensuring precise synchronization of the speech with the speaker's lip motion and adjusting the prosody to reflect facial emotion changes \cite{cong2024styledubber, chen2022v2c, zhang2024speaker, cong2024emodubber}.

In recent years, numerous outstanding works have been made in AVD.
By utilizing the speaker's lip motion from the current sentence, Neural Dubber \cite{hu2021neural}, DUS-AVO \cite{DSU-AVO}, and DubWise \cite{sahipjohn2024dubwise} significantly improve duration control and audio-visual synchronization. VDTTS \cite{VDTTS}, HPMDubbing \cite{HPMDubbing}, and 3D-VD \cite{3D-VD} enhance prosody by modeling current sentence's face frames. MCDubber \cite{zhao2024mcdubber} further enhances prosody expressiveness by temporally concatenating multimodal context.


Despite the progress, two key issues are overlooked: 1) Multiscale prosody expression attributes (including both sentence-level and phoneme-level) in the multimodal context influence the current sentence's prosody. 2) The prosody cues in the multimodal context should not be viewed in isolation, as their interaction with the current sentence influences the final prosody expressiveness. Previous studies in some related tasks, such as audiobook synthesis, have demonstrated the importance of multiscale modeling of multimodal context in enhancing speech expressiveness \cite{lei22c_interspeech, lei2023msstyletts}. For example, Chen et al. \cite{chen2022unsupervised} introduced a multiscale hierarchical context encoder, aided by a multiscale reference encoder, to predict both global and local context style embeddings, effectively improving the speech expressiveness. Inspired by these related works, how to model the multiscale multimodal context and facilitate its deep interaction with the current sentence to further enhance dubbing prosody expressiveness is the focus of this work.



\begin{figure*}[ht]
    \centering
    \includegraphics[width=\linewidth]{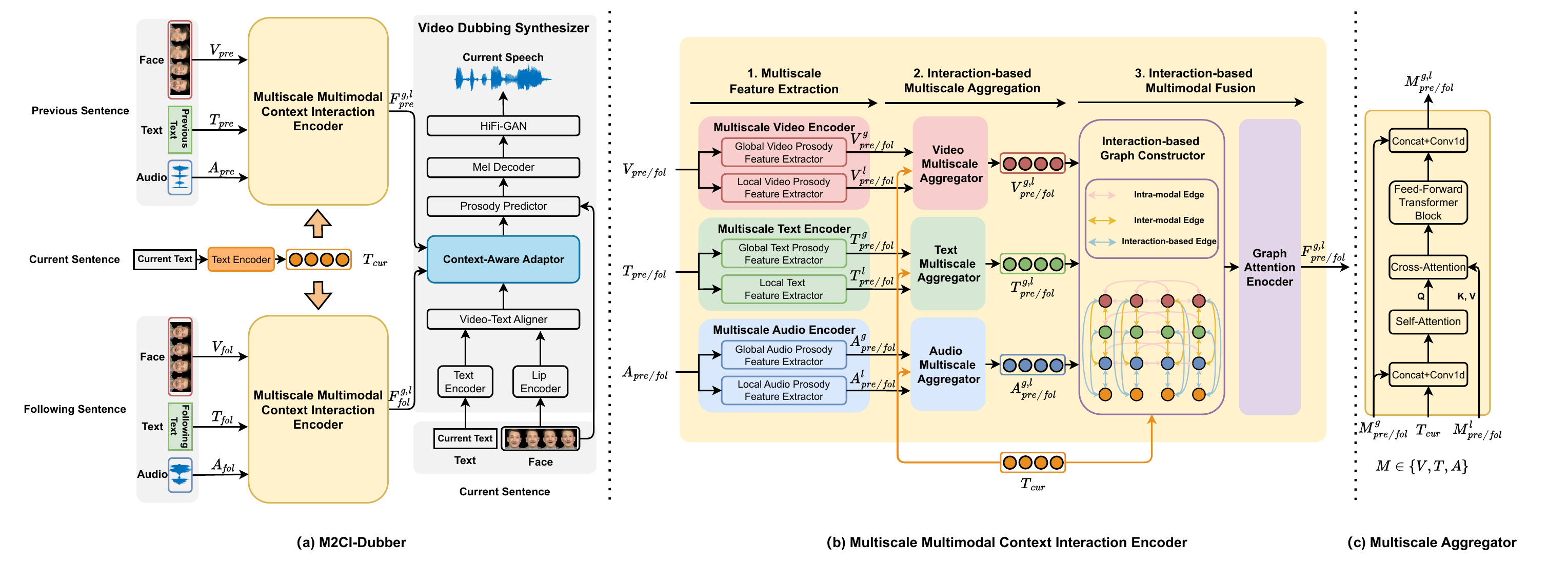}
    \vspace{-8mm}
    \caption{The proposed M2CI-Dubber contains two shared Multiscale Multimodal Context Interaction Encoders, a Text Encoder, and a Video Dubbing Synthesizer.}
    \label{fig_model}
\end{figure*}

To this end, we propose a Multiscale Multimodal Context Interaction (M2CI) scheme for AVD, termed \textbf{M2CI-Dubber}. Specifically, we design two shared M2CI encoders to model multiscale multimodal context and facilitate its deep interaction with the current sentence. The M2CI encoder consists of three key processes:
1) In Multiscale Feature Extraction (MFE), we utilize three Multiscale Encoders for each modality to generate global sentence-level and local phoneme-level features. 2) The Interaction-based Multiscale Aggregation (IMA) includes three multiscale aggregators for each modality. Within each aggregator, self-attention facilitates interaction between current text features and global features, followed by cross-attention to facilitate the interaction between the current text features and the local features. The aggregated output is further improved by incorporating global features through residual concatenation. 3) Interaction-based Multimodal Fusion (IMF) merges aggregated features from each modality using an interaction-based graph attention network with intra-modal and inter-modal edges. This network enables deep interaction by connecting current text features to nodes from each modality at the same steps simultaneously. The fused global-local features are then processed by a Context-Aware Adaptor in the Video Dubbing Synthesizer, which employs gated fusion and cross-attention to generate the current speech.
The main contributions of this paper are as follows:
\vspace{-1mm}
\begin{itemize}
    \item We propose a Multiscale Multimodal Context Interaction (M2CI) scheme for AVD, termed \textbf{M2CI-Dubber}, to model the multiscale multimodal context and facilitate its deep interaction with the current sentence.
    \item We design MFE to generate global and local features for each modality in the context. Additionally, we propose IMA to use attention-based mechanisms for aggregating the generated features while interacting with the current sentence. IMF is designed to employ an interaction-based graph attention network to fuse the aggregated features and interact with the current sentence.
    \item Experimental results on the Chem dataset show M2CI-Dubber outperforms baselines in dubbing expressiveness.
\end{itemize}

\vspace{-3mm}
\section{METHODOLOGY}
As shown in Fig. \ref{fig_model}(a), our M2CI-Dubber aims to generate speech for the current sentence given its text and face frames, as well as the face frames, text, and audio from the previous and following sentences. First, the current text is input into the Text Encoder to generate current text feature. Then, the face frames, text, and audio from the previous and following sentences are respectively fed into their own M2CI encoders, while interacting with the current text feature to generate fused global-local features. Finally, these fused global-local features from the previous and following sentences, along with the current sentence's text and face frames, are together sent to the Video Dubbing Synthesizer to generate current speech.

\subsection{Multiscale Feature Extraction }

\subsubsection{Multiscale Video Encoder}
In the Global Video Prosody Feature Extractor, face frames \(V_{\mathit{pre/fol}}\) from the previous or following sentence are processed by a dynamic facial expression recognition model MAE-DFER \cite{MAE-DFER}, fine-tuned on DFEW \cite{DFEW}, to generate the global sentence-level video prosody feature \(V_{\mathit{pre/fol}}^{\mathit{g}}\). In the Local Video Prosody Feature Extractor, face frames \(V_{\mathit{pre}} / V_{\mathit{fol}} \) are processed by an emotion face-alignment network \cite{emofan} to generate frame-level (with a temporal resolution of 40ms) video prosody feature. Inspired by the excellent work on multiscale style control in expressive speech synthesis \cite{quasi_phoneme_level}, the frame-level video prosody features are then fed into a video downsample encoder (where the temporal resolution is reduced 4 times) to generate the local quasi-phoneme-level video prosody feature \(V_{\mathit{pre/fol}}^{\mathit{l}} \). The video downsample encoder consists of two convolutional layers, Tanh activation, a linear layer, and a Feed-Forward Transformer (FFT) block \cite{fastspeech1}. Each convolutional layer has 256 filters, a $3 \times 1$ kernel, ReLU, batch normalization, dropout, and average pooling (kernel size 2).
\subsubsection{Multiscale Text Encoder}
In the Global Text Prosody Feature Extractor, the text \(T_{\mathit{pre / fol}} \) from the previous or following sentence is processed by a Roberta-based text emotion recognition model \cite{EERL} to generate the global sentence-level text prosody feature \(T_{\mathit{pre/fol}}^{\mathit{g}} \). In the Local Text Feature Extractor, text \(T_{\mathit{pre/fol}} \) is fed into a Text Encoder from the text-to-speech (TTS) model Fastspeech2 \cite{fastspeech2} to generate the local phoneme-level text feature \(T_{\mathit{pre/fol}}^{\mathit{l}} \).
\subsubsection{Multiscale Audio Encoder}
In Global Audio Prosody Feature Extractor, the audio \(A_{\mathit{pre/fol}}\) from the previous or following sentence is processed by Wav2Vec 2.0 \cite{baevski2020wav2vec}, fine-tuned on IEMOCAP \cite{busso2008iemocap}, to generate the global sentence-level audio prosody feature \(A_{\mathit{pre/fol}}^{\mathit{g}} \). In the Local Audio Prosody Feature Extractor, the audio \(A_{\mathit{pre}} / A_{\mathit{fol}}\) is processed by an emotion-aware speech self-supervised representation learning model \cite{liu2024ma_emotion} to generate frame-level (with a temporal resolution of 10ms) audio prosody features. The frame-level audio prosody features are then fed into an audio downsample encoder (where the temporal resolution is reduced 16 times) to generate the local quasi-phoneme-level audio prosody feature \(A_{\mathit{pre/fol}}^{\mathit{l}}\). The audio downsample encoder has the same structure as the video downsample encoder but includes four convolutional layers.
\subsection{Interaction-based 
Multiscale Aggregation }
As shown in the middle of Fig. \ref{fig_model}(b), we designed three multiscale aggregators with the same structure for each modality. Let \(M_{\mathit{pre/fol}}^{g}\) represent the global feature from the previous or following sentence, and \(M_{\mathit{pre/fol}}^{l}\) represent the local feature, where \(M \in \{V, T, A\}\). \(T_{\mathit{cur}}\) represents the current text feature, generated from the current sentence by a Text Encoder \cite{fastspeech2}. As shown in Fig. \ref{fig_model}(c), first, the current text feature \(T_{\mathit{cur}}\) is concatenated with the global feature \(M_{\mathit{pre/fol}}^{g}\) at each time step, followed by Conv1d to restore the original dimensions before concatenation. The output of Conv1d is then fed into a self-attention mechanism to enable interaction between the current text feature \(T_{\mathit{cur}}\) and the global feature \(M_{\mathit{pre/fol}}^{g}\). Next, cross-attention facilitates the interaction between the current text feature \(T_{\mathit{cur}}\) and the local feature \(M_{\mathit{pre/fol}}^{l}\), where the self-attention output acts as the query, and the local feature \(M_{\mathit{pre/fol}}^{l}\) serves as the key and value. The cross-attention output is then passed to the FFT block and concatenated with the global feature \(M_{\mathit{pre/fol}}^{g}\) at each time step to enhance the aggregated global-local feature output \(M_{\mathit{pre/fol}}^{g,l}\). In this way, the model not only aggregates global and local features but also enables  integration with the current text.
\subsection{Interaction-based  
Multimodal Fusion }
\subsubsection{Interaction-based Graph Constructor}
We introduce video global-local feature \(V_{\mathit{pre/fol}}^{\mathit{g,l}}\), text global-local feature \(T_{\mathit{pre/fol}}^{\mathit{g,l}}\) , the audio global-local feature \(A_{\mathit{pre/fol}}^{\mathit{g,l}}\) and current text feature \(T_{\mathit{cur}}\) to build an interaction-based undirected graph \(\mathcal{G} = (\mathcal{V}, \mathcal{E}) \), where \(\mathcal{V} \) is the set of nodes, and \(\mathcal{E} \) is the set of edges. For the edges of the graph, we define the following three types:  (1) \textbf{Intra-modal Edge} \( E_{\mathit{intra}} \) is generated between two nodes that are in same modalities but at different time steps.  (2) \textbf{Inter-modal Edge} \( E_{\mathit{inter}} \) is generated between two nodes that are in different modalities but at the same time steps. (3) \textbf{Interaction-based Edge} \( E_{\mathit{interaction}} \) is generated connecting current text feature to nodes from each modality at the same time step. Finally, the nodes are represented as \( \mathcal{V} = \{V_{\mathit{pre/fol}}^{\mathit{g,l}}, T_{\mathit{pre/fol}}^{\mathit{g,l}}, A_{\mathit{pre/fol}}^{\mathit{g,l}}, T_{\mathit{cur}}\} \), and the edges as \( \mathcal{E} = \{E_{\mathit{intra}}, E_{\mathit{inter}}, E_{\mathit{interaction}}\} \).

\subsubsection{Graph Nodes Fusion and Assessment} 
We feed \( \mathcal{V} \) and \( \mathcal{E} \) into a Graph Attention Encoder (GAE) for multimodal fusion, enabling interaction between the multiscale multimodal context and the current text. GAE consists of a multi-head graph attention layer \cite{GAT}, dropout, a linear layer, GELU activation, residual connections with the input, and layer normalization. The nodes in each modality incorporate information from the same modality, other modalities, and the current text feature. The fused nodes \( \{ \hat{V}_{\mathit{pre/fol}}^{\mathit{g,l}}, \hat{T}_{\mathit{pre/fol}}^{\mathit{g,l}}, \hat{A}_{\mathit{pre/fol}}^{\mathit{g,l}}, \hat{T}_{\mathit{cur}} \} \) are concatenated along the hidden dimension, passed through a Conv1d layer, and used as fused global-local output \( F_{\mathit{pre/fol}}^{\mathit{g,l}} \).
\subsection{Video Dubbing Synthesizer}
We use HPMDubbing \cite{HPMDubbing} as the backbone for Video Dubbing Synthesizer. The current text and lip frames (cropped from face frames) are input into the Text and Lip Encoders. Their outputs are fed into a Video-Text Aligner for duration control and audio-visual synchronization, producing $\mathit{H}_{\mathit{text, lip}}$. In the Context-Aware Adaptor (CAA), it considers that the influence of multiscale prosody expression attributes from different sentences to the current sentence may vary. The gated fusion results of $F_{\mathit{pre}}^{\mathit{g,l}}$ and $F_{\mathit{fol}}^{\mathit{g,l}}$ serve as the keys and values in cross-attention, with $H_{\mathit{text, lip}}$ as the query. A residual connection merges $\mathit{H}_{\mathit{text, lip}}$ with the cross-attention output, forming the CAA output. The output of CAA, along with the current face frames, is passed to the Prosody Predictor \cite{fastspeech2} for pitch and energy prediction. The output of Prosody Predictor is sent to the Mel Decoder to generate mel-spectrogram, which is then converted to the current speech by HiFi-GAN \cite{hifigan}.

\section{EXPERIMENTAL SETUP}

\begin{table*}[ht]
\setlength{\abovecaptionskip}{3pt} 
\setlength{\belowcaptionskip}{-3pt}
\caption{Objective and subjective (with 95\% confidence interval) evaluation results with other methods. \( \uparrow \) ($\downarrow$) indicates that a higher (lower) value is better, and bold indicates the best score. THE M2CI-Dubber SIGNIFICANTLY OUTPERFORM THE BASELINES WITH P-VALUE $<$ 0.001}
\label{table1_evaluations}
\centering

\renewcommand{\arraystretch}{1.2} 
\setlength{\tabcolsep}{5.5pt}
\resizebox{\textwidth}{!}{%

\begin{tabular}{lcccccccc}
\hline
Methods & GPE \( (\downarrow) \) & FFE \( (\downarrow) \) & LSE-C \( (\uparrow) \) & LSE-D \( (\downarrow) \) & WER \( (\downarrow) \) & MOS-C \( (\uparrow) \) & MOS-S \( (\uparrow) \) & AV Sync \( (\uparrow) \) \\
\hline
Ground-Truth & N/A & N/A & 8.722 & 6.564 & 03.85 & 4.490 ± 0.054 & 4.506 ± 0.047 & 4.501 ± 0.059 \\
Mel Resynthesis & 02.77 & 09.70 & 8.468 & 6.538 & 04.19 & 4.373 ± 0.056 & 4.370 ± 0.056 & 4.313 ± 0.060 \\
\hline
FastSpeech2 \cite{fastspeech2} & 50.26 & 41.24 & 3.438 & 9.399 & 13.66 & 3.325 ± 0.056 & 3.265 ± 0.059 & 3.161 ± 0.061 \\
DSU-AVO \cite{DSU-AVO} & 48.02 & 32.31 & \textbf{8.198} & 7.120 & \textbf{12.73} & 3.512 ± 0.050 & 3.469 ± 0.050 & 3.648 ± 0.038 \\
HPMDubbing \cite{HPMDubbing} & 42.78 & 36.56 & 7.934 & 7.057 & 18.66 & 3.752 ± 0.053 & 3.664 ± 0.044 & 3.612 ± 0.047 \\
MCDubber \cite{zhao2024mcdubber} & 40.26 & 30.94 & 7.571 & 6.943 & 17.96 & 3.774 ± 0.046 & 3.720 ± 0.046 & 3.740 ± 0.044 \\
\hline
\textbf{M2CI-Dubber (Proposed)} & \textbf{38.01} & \textbf{30.91} & 7.995 & \textbf{6.913} & 12.85 & \textbf{3.943 ± 0.044} & \textbf{3.866 ± 0.057} & \textbf{3.862 ± 0.053} \\
\hline

\end{tabular}%
}
\end{table*}

\begin{table}[ht]
\setlength{\abovecaptionskip}{3pt} 
\setlength{\belowcaptionskip}{-3pt}
\caption{Objective and subjective (with 95\% confidence interval) evaluation results of ablation studies.}
\label{table_ablation_studies}
\centering
\renewcommand{\arraystretch}{1.2} 
\setlength{\tabcolsep}{1.4pt} 
\resizebox{1.0\linewidth}{!}{
\begin{tabular}{p{0.31\linewidth}cccc} 
\hline
Methods & GPE \( (\downarrow) \) & FFE \( (\downarrow) \) & MOS-C \( (\uparrow) \) & MOS-S \( (\uparrow) \) \\
\hline
w/o Global model & 42.63 & 31.12 & 3.788 ± 0.040 & 3.751 ± 0.045 \\
w/o Local model & 44.97 & 31.96 & 3.705 ± 0.044 & 3.682 ± 0.045 \\
w/o IMA & 41.81 & 31.72 & 3.780 ± 0.047 & 3.767 ± 0.047 \\
w/o IMF & 43.94 & 31.95 & 3.742 ± 0.043 & 3.721 ± 0.055 \\
w/o CAA & 40.97 & 31.10 & 3.801 ± 0.051 & 3.730 ± 0.045 \\
\hline
w/o Int. in IMA & 40.59 & 31.79 & 3.769 ± 0.039 & 3.732 ± 0.056 \\
w/o Int. in IMF & 42.40 & 30.98 & 3.752 ± 0.040 & 3.757 ± 0.045 \\
w/o Int. in IMA \(\&\)IMF & 41.32 & 31.18 & 3.719 ± 0.049 & 3.685 ± 0.051 \\
\hline
w/o Video & 41.36 & 31.07 & 3.767 ± 0.045 & 3.742 ± 0.051 \\
w/o Text & 42.02 & 31.21 & 3.771 ± 0.044 & 3.733 ± 0.045 \\
w/o Audio & 41.41 & 31.55 & 3.738 ± 0.048 & 3.716 ± 0.048 \\
\hline
w/o Previous Sentence & 43.26 & 31.39 & 3.725 ± 0.046 & 3.717 ± 0.047 \\
w/o Following Sentence & 42.65 & 31.04 & 3.727 ± 0.043 & 3.748 ± 0.049 \\
\hline
\textbf{M2CI-Dubber}  & \textbf{38.01} & \textbf{30.91} & \textbf{3.943 ± 0.044} & \textbf{3.866 ± 0.057} \\
\hline

\end{tabular}%
}
\end{table}

\subsection{Dataset}
The Chem dataset is a single-speaker English speech dataset consisting of video clips and corresponding transcripts, featuring rich speech prosody [20]. During the original collection process, clips without visible faces were filtered out, resulting in non-consecutive samples in the original dataset. Therefore, previous work re-collected consecutive samples to create the Context Chem dataset [7], with the training set containing 3,308 consecutive samples, the validation set with 85 samples, and the test set with 113 samples. We will evaluate our model on the Context Chem dataset.
\vspace{-2mm}
\subsection{Implementation Details}

The global and local features of each modality and current text feature are set to 256 dimensions. And, we use two attention heads for GAE, and the fused features are also set to 256 dimensions. Each video clip is processed at 25 FPS with a sampling rate of 16000 Hz. Mel-spectrogram features are extracted with a 40ms window length and 10ms hop length. We use Adam optimizer with an initial learning rate of 0.00625, $\beta_1 = 0.9$, $\beta_2 = 0.98$, and $\epsilon = 10^{-9}$, with batch size is 8 and 40k training steps. Our model is trained on A800 80G GPU. The Video Dubbing Synthesizer was pre-trained following the same train process as HPMDubbing \cite{HPMDubbing}. \subsection{Evaluation Metrics}
\subsubsection{Objective metrics}
\vspace{-1mm}
(1) Gross Pitch Error (GPE) \cite{GPE}: percentage of frames where pitch error exceeds 20 \% with voicing present in both synthesized and ground-truth speech.
(2) F0 Frame Error (FFE) \cite{FFE}: percentage of frames with either a voicing decision error or a pitch error exceeding 20\%. 
(3) Lip Sync Error-Confidence (LSE-C) \cite{syncnet, lipexpert}:  average confidence score, where higher scores indicate better audio-video correlation.
(4) Lip Sync Error-Distance (LSE-D) \cite{syncnet, lipexpert}: average error by measuring the distance between lip and audio representations.
(5) Word Error Rate (WER): measures pronunciation accuracy using the whisper-large-v3 automatic speech recognition model \cite{whisper}.
\subsubsection{Subjective metrics}
We conducted a Mean Opinion Score (MOS) \cite{liu2024text, he2024multi, liu2024multi, liu2024emphasis, liu2024generative, hu2024fctalker} test with 20 trained raters who evaluated 12 generated dubbed videos and speech samples, scoring the following metrics on a scale of 1 to 5:
(1) MOS-Context (MOS-C): The dubbed video is spliced with the original context video to evaluate how well the prosody aligns with the multimodal context.
(2) MOS-Similarity (MOS-S): Evaluates the similarity of the synthesized speech's prosody to the ground-truth.
(3) Audio-Visual Synchronization (AV Syn) \cite{hu2021neural}: Assesses AV synchronization by overlaying synthesized speech onto video.

\section{RESULTS AND DISCUSSION}

\subsection{Main Results}

We developed four baseline models as follows. (1) \textbf{FastSpeech2} \cite{fastspeech2}: a neural TTS model; (2) \textbf{DSU-AVO} \cite{DSU-AVO}, an AVD model whose learning objective is discrete speech unit prediction; (3) \textbf{HPMDubbing} \cite{HPMDubbing}, an AVD model incorporating hierarchical prosody modeling by considering the current sentence's face frames. (4) \textbf{MCDubber} \cite{zhao2024mcdubber}, an AVD model incorporating temporally concatenating the multimodal context. As shown in Table \ref{table1_evaluations}, M2CI-Dubber achieved the best results across all prosody-related metrics. Additionally, statistical analysis indicates that M2CI-Dubber significantly outperforms the baselines with a p-value less than 0.001.  
For objective metrics, M2CI-Dubber achieved the best scores in GPE (38.01) and FFE (30.91), demonstrating that M2CI-Dubber is able to generate speech with prosody closer to the Ground Truth compared to other baselines. In terms of subjective metrics, our model also achieved the highest scores in MOS-C (3.943) and MOS-S (3.866), indicating that M2CI-Dubber can better enhance the prosody expressiveness of the dubbed video and align its prosody with context.
\subsection{Ablation Results}
We conducted a comprehensive ablation study as follows:  
(1) \textbf{Component Ablation}: We removed the following components: Multiscale Feature Extraction (w/o Global model, w/o Local model), Interaction-based Multiscale Aggregation (w/o IMA), Interaction-based Multimodal Fusion (w/o IMF), and Context-Aware Adaptor (w/o CAA), as shown in Rows 2-6 of Table \ref{table_ablation_studies}. Removing these components increased GPE and FFE while reducing MOS-C and MOS-S, demonstrating the contribution of each component for prosody expressiveness and alignment of prosody with the multimodal context. 
(2) \textbf{Interaction Ablation}: We designed three interaction-related ablation models by removing interaction with the current text in IMA (w/o Int. in IMA), IMF (w/o Int. in IMF), and both (w/o Int. in IMA \(\&\)IMF). The results, shown in Rows 7-9 of Table \ref{table_ablation_studies}, indicate that the largest decline in subjective metrics occurred when interaction is removed both both in IMA and IMF. This highlights the crucial role of maintaining deep interaction in aggregation and fusion process for better prosody generation.
(3) \textbf{Modality Ablation}: The modality ablation study, where video, text, and audio were respectively removed (Rows 10-12 of Table \ref{table_ablation_studies}), resulted in higher GPE and FFE, and lower MOS-C and MOS-S. These results indicate that multiscale modeling of each modality context is essential for dubbing prosody. (4) \textbf{Context Ablation}: Lastly, we removed the multimodal input from the previous and following sentences, respectively (Rows 13-14 of Table \ref{table_ablation_studies}). This led to a decline in the performance of prosody-related metrics, demonstrating that multiscale modeling of both previous and following sentences’ multimodal context is crucial for improving prosody.
\vspace{-2mm}
\section{CONCLUSION}
\vspace{-2mm}
In this paper, we propose \textbf{M2CI-Dubber}, a Multiscale Multimodal Context Interaction scheme for expressive video dubbing, to model multiscale multimodal context and facilitate its deep interaction with the current sentence. In future work, we will explore the influence of multiscale multimodal context on modeling emotion expressiveness in AVD.
\vspace{-1mm}
\section{ACKNOWLEDGMENTS}
\vspace{-1mm}
This work was funded by the Young Scientists Fund (No. 62206136) and the General Program (No. 62476146) of the National Natural Science Foundation of China, the ``Inner Mongolia Science and Technology Achievement Transfer and Transformation Demonstration Zone, University Collaborative Innovation Base, and University Entrepreneurship Training Base'' Construction Project (Supercomputing Power Project) (No.21300-231510).


\bibliographystyle{IEEEtran}
\bibliography{ref_unsort}

\begin{thebibliography}{10}
\providecommand{\url}[1]{#1}
\csname url@samestyle\endcsname
\providecommand{\newblock}{\relax}
\providecommand{\bibinfo}[2]{#2}
\providecommand{\BIBentrySTDinterwordspacing}{\spaceskip=0pt\relax}
\providecommand{\BIBentryALTinterwordstretchfactor}{4}
\providecommand{\BIBentryALTinterwordspacing}{\spaceskip=\fontdimen2\font plus
\BIBentryALTinterwordstretchfactor\fontdimen3\font minus \fontdimen4\font\relax}
\providecommand{\BIBforeignlanguage}[2]{{%
\expandafter\ifx\csname l@#1\endcsname\relax
\typeout{** WARNING: IEEEtran.bst: No hyphenation pattern has been}%
\typeout{** loaded for the language `#1'. Using the pattern for}%
\typeout{** the default language instead.}%
\else
\language=\csname l@#1\endcsname
\fi
#2}}
\providecommand{\BIBdecl}{\relax}
\BIBdecl

\bibitem{hu2021neural}
C.~Hu, Q.~Tian, T.~Li, W.~Yuping, Y.~Wang, and H.~Zhao, ``Neural dubber: Dubbing for videos according to scripts,'' \emph{Advances in neural information processing systems}, vol.~34, pp. 16\,582--16\,595, 2021.

\bibitem{3D-VD}
Z.~Yang, S.~Liu, X.~Li, H.~Wu, Z.~Wu, Y.~Shan, and J.~Jia, ``{Prosody Modeling with 3D Visual Information for Expressive Video Dubbing},'' in \emph{Proc. INTERSPEECH 2023}, 2023, pp. 4863--4867.

\bibitem{cong2024styledubber}
G.~Cong, Y.~Qi, L.~Li, A.~Beheshti, Z.~Zhang, A.~v.~d. Hengel, M.-H. Yang, C.~Yan, and Q.~Huang, ``Styledubber: Towards multi-scale style learning for movie dubbing,'' \emph{arXiv preprint arXiv:2402.12636}, 2024.

\bibitem{chen2022v2c}
Q.~Chen, M.~Tan, Y.~Qi, J.~Zhou, Y.~Li, and Q.~Wu, ``V2c: Visual voice cloning,'' in \emph{Proceedings of the IEEE/CVF Conference on Computer Vision and Pattern Recognition}, 2022, pp. 21\,242--21\,251.

\bibitem{zhang2024speaker}
Z.~Zhang, L.~Li, G.~Cong, Y.~Haibing, Y.~Gao, C.~Yan, A.~van~den Hengel, and Y.~Qi, ``From speaker to dubber: Movie dubbing with prosody and duration consistency learning,'' in \emph{ACM Multimedia 2024}.

\bibitem{cong2024emodubber}
G.~Cong, J.~Pan, L.~Li, Y.~Qi, Y.~Peng, A.~v.~d. Hengel, J.~Yang, and Q.~Huang, ``Emodubber: Towards high quality and emotion controllable movie dubbing,'' \emph{arXiv preprint arXiv:2412.08988}, 2024.

\bibitem{DSU-AVO}
J.~Lu, B.~Sisman, M.~Zhang, and H.~Li, ``High-quality automatic voice over with accurate alignment: Supervision through self-supervised discrete speech units,'' \emph{arXiv preprint arXiv:2306.17005}, 2023.

\bibitem{sahipjohn2024dubwise}
N.~Sahipjohn, A.~Gudmalwar, N.~Shah, P.~Wasnik, and R.~R. Shah, ``Dubwise: Video-guided speech duration control in multimodal llm-based text-to-speech for dubbing,'' \emph{arXiv preprint arXiv:2406.08802}, 2024.

\bibitem{VDTTS}
M.~Hassid, M.~T. Ramanovich, B.~Shillingford, M.~Wang, Y.~Jia, and T.~Remez, ``More than words: In-the-wild visually-driven prosody for text-to-speech,'' in \emph{Proceedings of the IEEE/CVF Conference on Computer Vision and Pattern Recognition}, 2022, pp. 10\,587--10\,597.

\bibitem{HPMDubbing}
G.~Cong, L.~Li, Y.~Qi, Z.-J. Zha, Q.~Wu, W.~Wang, B.~Jiang, M.-H. Yang, and Q.~Huang, ``Learning to dub movies via hierarchical prosody models,'' in \emph{Proceedings of the IEEE/CVF Conference on Computer Vision and Pattern Recognition}, 2023, pp. 14\,687--14\,697.

\bibitem{zhao2024mcdubber}
Y.~Zhao, Z.~Jia, R.~Liu, D.~Hu, F.~Bao, and G.~Gao, ``Mcdubber: Multimodal context-aware expressive video dubbing,'' \emph{arXiv preprint arXiv:2408.11593}, 2024.

\bibitem{lei22c_interspeech}
S.~Lei, Y.~Zhou, L.~Chen, J.~Hu, Z.~Wu, S.~Kang, and H.~Meng, ``{Towards Multi-Scale Speaking Style Modelling with Hierarchical Context Information for Mandarin Speech Synthesis},'' in \emph{Proc. Interspeech 2022}, 2022, pp. 5523--5527.

\bibitem{lei2023msstyletts}
S.~Lei, Y.~Zhou, L.~Chen, Z.~Wu, X.~Wu, S.~Kang, and H.~Meng, ``Msstyletts: Multi-scale style modeling with hierarchical context information for expressive speech synthesis,'' \emph{IEEE/ACM Transactions on Audio, Speech, and Language Processing}, 2023.

\bibitem{chen2022unsupervised}
X.~Chen, S.~Lei, Z.~Wu, D.~Xu, W.~Zhao, and H.~Meng, ``Unsupervised multi-scale expressive speaking style modeling with hierarchical context information for audiobook speech synthesis,'' in \emph{Proceedings of the 29th International Conference on Computational Linguistics}, 2022, pp. 7193--7202.

\bibitem{MAE-DFER}
L.~Sun, Z.~Lian, B.~Liu, and J.~Tao, ``Mae-dfer: Efficient masked autoencoder for self-supervised dynamic facial expression recognition,'' in \emph{Proceedings of the 31st ACM International Conference on Multimedia}, 2023, pp. 6110--6121.

\bibitem{DFEW}
X.~Jiang, Y.~Zong, W.~Zheng, C.~Tang, W.~Xia, C.~Lu, and J.~Liu, ``Dfew: A large-scale database for recognizing dynamic facial expressions in the wild,'' in \emph{Proceedings of the 28th ACM international conference on multimedia}, 2020, pp. 2881--2889.

\bibitem{emofan}
A.~Toisoul, J.~Kossaifi, A.~Bulat, G.~Tzimiropoulos, and M.~Pantic, ``Estimation of continuous valence and arousal levels from faces in naturalistic conditions,'' \emph{Nature Machine Intelligence}, vol.~3, no.~1, pp. 42--50, 2021.

\bibitem{quasi_phoneme_level}
X.~Li, C.~Song, J.~Li, Z.~Wu, J.~Jia, and H.~Meng, ``Towards multi-scale style control for expressive speech synthesis,'' \emph{arXiv preprint arXiv:2104.03521}, 2021.

\bibitem{fastspeech1}
Y.~Ren, Y.~Ruan, X.~Tan, T.~Qin, S.~Zhao, Z.~Zhao, and T.-Y. Liu, ``Fastspeech: Fast, robust and controllable text to speech,'' \emph{Advances in neural information processing systems}, vol.~32, 2019.

\bibitem{EERL}
J.~Hartmann, ``Emotion-english-roberta-large,'' \url{https://huggingface.co/j-hartmann/emotion-english-roberta-large}, 2021.

\bibitem{fastspeech2}
Y.~Ren, C.~Hu, X.~Tan, T.~Qin, S.~Zhao, Z.~Zhao, and T.-Y. Liu, ``Fastspeech 2: Fast and high-quality end-to-end text to speech,'' \emph{arXiv preprint arXiv:2006.04558}, 2020.

\bibitem{baevski2020wav2vec}
A.~Baevski, Y.~Zhou, A.~Mohamed, and M.~Auli, ``wav2vec 2.0: A framework for self-supervised learning of speech representations,'' \emph{Advances in neural information processing systems}, vol.~33, pp. 12\,449--12\,460, 2020.

\bibitem{busso2008iemocap}
C.~Busso, M.~Bulut, C.-C. Lee, A.~Kazemzadeh, E.~Mower, S.~Kim, J.~N. Chang, S.~Lee, and S.~S. Narayanan, ``Iemocap: Interactive emotional dyadic motion capture database,'' \emph{Language resources and evaluation}, vol.~42, pp. 335--359, 2008.

\bibitem{liu2024ma_emotion}
R.~Liu and Z.~Ma, ``Emotion-aware speech self-supervised representation learning with intensity knowledge,'' \emph{arXiv preprint arXiv:2406.06646}, 2024.

\bibitem{GAT}
P.~Veli{\v{c}}kovi{\'c}, G.~Cucurull, A.~Casanova, A.~Romero, P.~Lio, and Y.~Bengio, ``Graph attention networks,'' \emph{arXiv preprint arXiv:1710.10903}, 2017.

\bibitem{hifigan}
J.~Kong, J.~Kim, and J.~Bae, ``Hifi-gan: Generative adversarial networks for efficient and high fidelity speech synthesis,'' \emph{Advances in neural information processing systems}, vol.~33, pp. 17\,022--17\,033, 2020.

\bibitem{GPE}
T.~Nakatani, S.~Amano, T.~Irino, K.~Ishizuka, and T.~Kondo, ``A method for fundamental frequency estimation and voicing decision: Application to infant utterances recorded in real acoustical environments,'' \emph{Speech Communication}, vol.~50, no.~3, pp. 203--214, 2008.

\bibitem{FFE}
W.~Chu and A.~Alwan, ``Reducing f0 frame error of f0 tracking algorithms under noisy conditions with an unvoiced/voiced classification frontend,'' in \emph{2009 IEEE International Conference on Acoustics, Speech and Signal Processing}.\hskip 1em plus 0.5em minus 0.4em\relax IEEE, 2009, pp. 3969--3972.

\bibitem{syncnet}
J.~S. Chung and A.~Zisserman, ``Out of time: automated lip sync in the wild,'' in \emph{Computer Vision--ACCV 2016 Workshops: ACCV 2016 International Workshops, Taipei, Taiwan, November 20-24, 2016, Revised Selected Papers, Part II 13}.\hskip 1em plus 0.5em minus 0.4em\relax Springer, 2017, pp. 251--263.

\bibitem{lipexpert}
K.~Prajwal, R.~Mukhopadhyay, V.~P. Namboodiri, and C.~Jawahar, ``A lip sync expert is all you need for speech to lip generation in the wild,'' in \emph{Proceedings of the 28th ACM international conference on multimedia}, 2020, pp. 484--492.

\bibitem{whisper}
A.~Radford, J.~W. Kim, T.~Xu, G.~Brockman, C.~McLeavey, and I.~Sutskever, ``Robust speech recognition via large-scale weak supervision,'' in \emph{International conference on machine learning}.\hskip 1em plus 0.5em minus 0.4em\relax PMLR, 2023, pp. 28\,492--28\,518.

\bibitem{liu2024text}
R.~Liu, Y.~Hu, H.~Zuo, Z.~Luo, L.~Wang, and G.~Gao, ``Text-to-speech for low-resource agglutinative language with morphology-aware language model pre-training,'' \emph{IEEE/ACM Transactions on Audio, Speech, and Language Processing}, 2024.

\bibitem{he2024multi}
S.~He, R.~Liu, and H.~Li, ``Multi-source spatial knowledge understanding for immersive visual text-to-speech,'' \emph{arXiv preprint arXiv:2410.14101}, 2024.

\bibitem{liu2024multi}
R.~Liu, S.~He, Y.~Hu, and H.~Li, ``Multi-modal and multi-scale spatial environment understanding for immersive visual text-to-speech,'' \emph{arXiv preprint arXiv:2412.11409}, 2024.

\bibitem{liu2024emphasis}
R.~Liu, Z.~Jia, J.~Yang, Y.~Hu, and H.~Li, ``Emphasis rendering for conversational text-to-speech with multi-modal multi-scale context modeling,'' \emph{arXiv preprint arXiv:2410.09524}, 2024.

\bibitem{liu2024generative}
R.~Liu, Y.~Hu, Y.~Ren, X.~Yin, and H.~Li, ``Generative expressive conversational speech synthesis,'' in \emph{Proceedings of the 32nd ACM International Conference on Multimedia}, 2024, pp. 4187--4196.

\bibitem{hu2024fctalker}
Y.~Hu, R.~Liu, G.~Gao, and H.~Li, ``Fctalker: Fine and coarse grained context modeling for expressive conversational speech synthesis,'' in \emph{2024 IEEE 14th International Symposium on Chinese Spoken Language Processing (ISCSLP)}.\hskip 1em plus 0.5em minus 0.4em\relax IEEE, 2024, pp. 299--303.

\end{thebibliography}

\end{document}